\documentclass[twocolumn,pra]{revtex4-1}

\usepackage{graphicx}
\usepackage{eso-pic}
\usepackage{color}
\usepackage{type1cm}
\usepackage{amssymb, amsmath}
\usepackage{subfigure}

\begin{document} 

\title{Novel Collective Effects in Integrated Photonics}
\author{ M. Delanty} 
\affiliation{Centre for Engineered Quantum Systems, Department of Physics and Astronomy, Macquarie University, Sydney, NSW 2109, Australia}
\affiliation{CSIRO Materials Science and Engineering, P.O. Box 218, Lindfield, NSW 2070, Australia}
\author{S. Rebi\'c}
\affiliation{Centre for Engineered Quantum Systems, Department Physics \& Astronomy, Macquarie University, North Ryde, NSW 2109, Australia}
\affiliation{Centre for Quantum Computation and Communication Technology, Department of Quantum Science, Australian National University, Canberra ACT 0200, Australia}\author{J. Twamley}
\affiliation{Centre for Engineered Quantum Systems, Department Physics \& Astronomy, Macquarie University, North Ryde, NSW 2109, Australia}

\begin{abstract}
Superradiance, the enhanced collective emission of energy from a coherent ensemble of quantum systems, has been typically studied in atomic ensembles. In this work we study theoretically the enhanced emission of energy from coherent ensembles of harmonic oscillators. We show that it should be possible to observe harmonic oscillator superradiance for the first time in waveguide arrays in integrated photonics. Furthermore, we describe how pairwise correlations within the ensemble can be measured with this architecture. These pairwise correlations are an integral part of the phenomenon of superradiance and have never been observed in experiments to date.
\end{abstract}

\maketitle

\section{Introduction}

Integrated photonics is a rapidly emerging experimental platform for the observation of novel quantum effects. With the advent of the direct write technique \cite{MarshallDirectWrite,DirectWrite} and the growing number of integrated `on-chip' optical components \cite{ObrienTuneableCoupler,ObrienMultimodeExperiment,ObrienNatureReview} it is now possible to build complex 2D and 3D waveguide structures to engineer a variety of multimode interactions. Recent experiments in integrated photonic devices have exhibited correlations in continuous time quantum walks \cite{IPRandomWalks,DirectWriteThorne}, a classical analogue of displaced Fock states \cite{SzameitGlauberFock} and an on chip correlated photon-pair source \cite{MikeSPDC}, to name but a few. However, despite the ability of the direct write technique to laser write waveguides in quite arbitrary 3D geometries, there has been little experimental work on intrinsically 3D waveguide arrays.

In this paper we show that it should be possible to observe novel quantum collective effects in a 3D integrated photonic device. In particular we theoretically show that an integrated photonic implementation could provide the first experimental observation of the phenomenon of harmonic oscillator superradiance. Atomic superradiance has been theoretically and experimentaly studied since 1954 when Dicke \cite{Dicke} showed that by confining an ensemble of two level atoms to a region that is small compared to the wavelength, particular states of the ensemble could radiate with an intensity proportional to the square of the number of atoms. This is in contrast to normal radiance where the intensity of a sparse ensemble of two level atoms is proportional to the number of atoms. This superradiance occurs due to the fact that a dense ensemble of two level atoms interacts with a common reservoir, rather than an ensemble of individual reservoirs. The reservoir can no longer distinguish which atom decayed and quantum interference between the many different decay pathways can occur. This interference can enhance or reduce the emission intensity from the ensemble leading to superradiance and subradiance respectively.

Following the work of Dicke, it was theoretically shown that superradiance can occur in a variety of systems including multilevel atoms and harmonic oscillators when the ensemble decays into a common reservoir \cite{Agarwal1974,HarocheReview}. There have also been many experimental and theoretical works studying various aspects of superradiant phenomena including correlations, pulse propagation within the ensemble, collisional dephasing of the gas molecules and polarization effects \cite{HarocheReview,BrandesReview}. Of particular note is the two-time correlations within the ensemble in superradiance which have been studied extensively theoretically \cite{Agarwal1974,AgarwalSRpaper,SRCorrelationsTheory,TwoIonSRTHeory,CarmichaelStatisticalMethods1}, however have not been experimentally measured due to the difficulty in isolating two atoms in a strongly confined ensemble.

Furthermore, despite over thirty years of superradiance experiments, harmonic oscillator superradiance has never been experimentally observed. This is due to the difficulty of engineering a common reservoir interaction for an ensemble of harmonic oscillators. As a result, there has been very little theoretical work on harmonic oscillator superradiance since the 1970's, when many properties of harmonic oscillator superradiance were calculated  \cite{PuriLawande,Agarwal1974,AgarwalQHOpaper}. Of particular relevance to this work are the general solutions for the superradiant intensity and two-time correlations derived by Agarwal \cite{Agarwal1974,AgarwalQHOpaper}. As these general expressions are complex, we find simpler solutions for specific states that are experimentally relevant to our integrated photonics implementation.

In this paper we show that waveguide arrays of optical modes (harmonic oscillators) each guided within its own 3D waveguide in integrated photonics can be used to engineer a common reservoir interaction for an  ensemble of waveguides. We demonstrate that harmonic oscillator (photonic) superradiance should be readily observed for the first time using existing integrated 3D waveguide technology. Furthermore, we find accurate analytic approximations for experimental observables that one can use to probe the photonic superradiance. We also propose a feasible experiment for measuring pairwise correlations within the waveguide ensemble. This proposal is significant as pairwise correlations have never before been measured in any superradiance experiment.

This paper is structured as follows. In section \ref{Section_IPSystem_Intro} we describe our integrated photonics proposal, whereby an ensemble of optical modes collectively decay into a common bath. The latter is simulated by a semi-infinite waveguide array. Furthermore we show how this implementation is well approximated by a photonic  harmonic oscillator version of the superradiance master equation. In section \ref{AnalyticResults} we theoretically calculate the intensity of light coupling into the bath waveguide array and two-time correlation functions between two waveguides in the ensemble. We also suggest an experimental method in the integrated photonics architecture to measure  the intensity and correlations of the photonic harmonic oscillator superradiance. These results are compared to a coupled mode theory simulation of the entire waveguide system in section \ref{NumericResults}. We conclude in section \ref{Conclusion}.

\section{An Integrated Photonics Implementation of the Superradiance Master Equation  \label{Section_IPSystem_Intro}  }

\begin{figure}
\begin{center}
\includegraphics[scale=0.275]{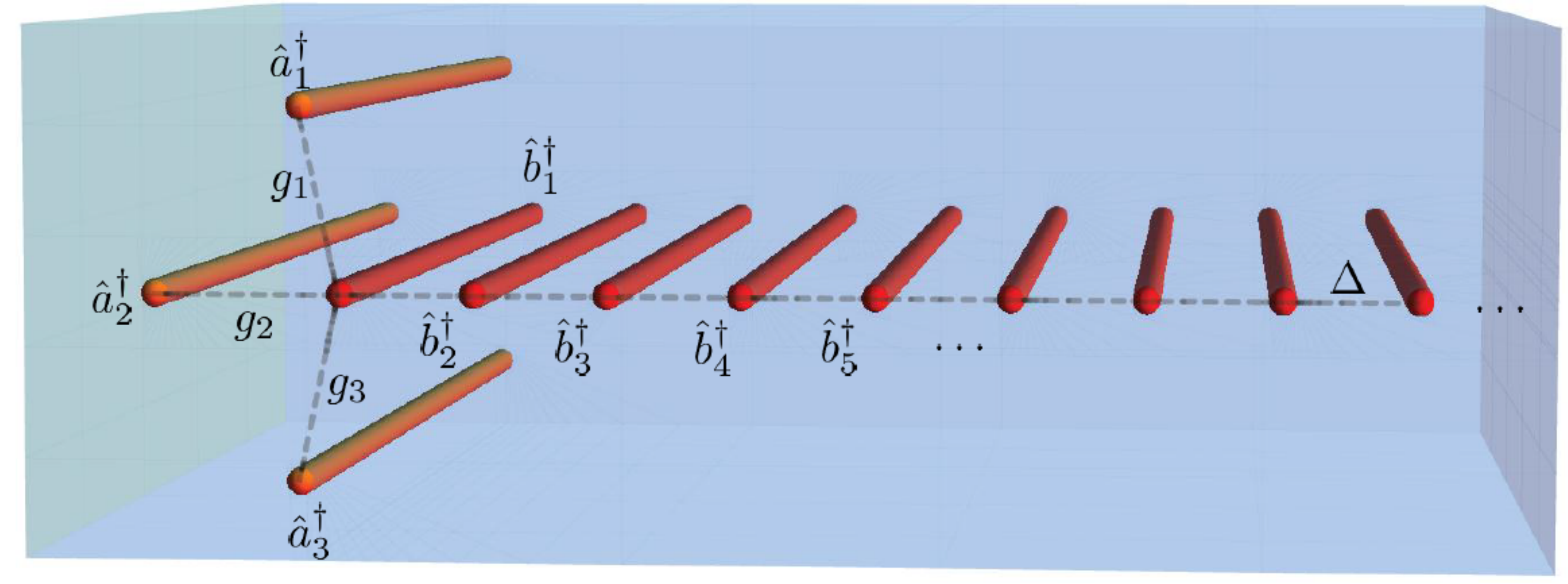}
\end{center}
\caption{ Waveguide geometry for the observation of superradiance. Light injected into the system waveguides, $\hat{a}_j$, will irreversibly decay into a semi-infinite array of bath waveguides, $\hat{b}_k$.}
\label{IP_Diagram}
\end{figure}

In this section we introduce the master equation for photonic harmonic oscillator superradiance. We then describe a 3D waveguide array in integrated photonics and show under certain conditions that the photonic modes propagating in the waveguide array can be described by the superradiance master equation. The superradiance master equation for oscillators describes the interaction of an ensemble of $N$ resonant, isolated oscillators with a common zero temperature heat bath. In the frame rotating at $\omega$ the superradiance master equation is \cite{Agarwal1974},
\begin{eqnarray}
\label{General_SRME}
\! \! \! \! \! \! \dot{\rho} &=&  \sum^N_{i,j=1} \frac{\gamma_{i,j}}{2} \left( 2 \hat{a}_j \rho \hat{a}_i^{\dagger} -  \hat{a}_i^{\dagger} \hat{a}_j \rho - \rho \hat{a}_i^{\dagger} \hat{a}_j  \right),
\end{eqnarray}
where, $\omega$ is the frequency of each oscillator and $\gamma_{i,j}$ is the rate of  decay of the $i$-th and $j$-th oscillators to the common bath. The dynamics of the superradiance master equation (\ref{General_SRME}) are significantly different from those of normal radiance where the oscillators have individual baths ($\gamma_{i\neq j}=0$) and decay independently of one another. Depending on the initial state the additional, $i\neq j$, terms in (\ref{General_SRME}) can enhance or reduce the decay of the ensemble, leading to superradiance or subradiance, respectively. Therefore the emission intensity of the ensemble into the bath can be varied simply by changing the symmetry of the initial state of the ensemble. This is in contrast to normal radiance, where the intensity can only be varied by increasing or reducing the amount of energy in the ensemble.

We now describe how the superradiance master equation can be engineered in an integrated photonics implementation. In particular we show how a common reservoir can be  engineered for an ensemble of waveguides by coupling the ensemble to a semi-infinite array of bath waveguides.  This extends recent work on reservoir engineering \cite{LonghiNonexp,chinesePRA,chinesepaper}, which considered only  a single system waveguide coupled to a semi-infinite bath waveguide array. Our proposal provides an example of how global reservoirs can be engineered for an ensemble of quantum systems, rather than the typical case engineering a reservoir for a single quantum system \cite{ZollerNAMRCPB}.

We consider an ensemble of $N$ system waveguides coupled to a semi-infinite linear array of bath waveguides  as depicted in Figure  \ref{IP_Diagram}.  We go beyond the tight binding limit commonly used in waveguide arrays \cite{LonghiNonexp,LonghiTightBindingEx} and write the full quantum Hamiltonian of the system  in the interaction picture as,
\begin{eqnarray} 
\label{Ham_AllCouplings}
\hat{H} &=& 
\sum^N_{j=1} g_j \left(\hat{a}^{\dagger}_j \hat{b}_1 + \hat{b}^{\dagger}_1 \hat{a}_j \right) +
\Delta \sum^{\infty}_{k=1} \left( \hat{b}^{\dagger}_k \hat{b}_{k+1} + \hat{b}^{\dagger}_{k+1} \hat{b}_{k} \right) \nonumber \\
&+& \sum^N_{i \neq j=1} \Omega_{i,j} \hat{a}^{\dagger}_i \hat{a}_j + \sum^N_{j=1} \left( \hat{a}^{\dagger}_j \hat{B}_j + \hat{B}^{\dagger}_j \hat{a}_j \right) 
\end{eqnarray}
where, the annihilation operators for the system  and bath waveguides are, $ \hat{a}_j$ and $ \hat{b}_k$, respectively, the $j$-th system waveguide couples to the first bath mode at the rate $g_j$, the bath waveguides couple to their nearest neighbours at the rate $\Delta$ and the $i$-th and $j$-th system waveguides couple at the rate $\Omega_{i,j}=\Omega_{j,i}$, which we will see later can be ignored under certain assumptions. Furthermore, the coupling of the $j$-th system waveguides to all bath waveguides, other than the first,  is described by the collective operator, $\hat{B}_j = \sum^{\infty}_{k=2}  J^{(j)}_k \hat{b}_k$, where $J^{(j)}_k$ is the rate of coupling between the $j$-th system waveguide and the $k$-th bath waveguide.

In order to engineer the superradiance master equation, we make several assumptions. The validity of these assumptions will be assessed through numerical simulations in section \ref{NumericResults}. First we assume that the system waveguides do not interact with one another, $\Omega_{i,j} \approx 0$, as these interactions unnecessarily complicate the dynamics. This can be achieved by maximizing the distance between the system waveguides, although this is difficult to do with a large number of waveguides. We next assume that the system waveguides only couple to the first bath waveguide, $J^{(j)}_k \approx 0$. This assumption greatly simplifies the calculation of the spectral density of the reservoir \cite{chinesepaper}. After making these two assumptions (\ref{Ham_AllCouplings}) becomes,
\begin{eqnarray}
\label{HIPwithBath}
\!\!\!\! \hat{H} &=& 
\mathcal{G}_N \left(\hat{C}^{\dagger}_N \hat{b}_1 + \hat{b}^{\dagger}_1 \hat{C}_N \right) +
\Delta \sum^{\infty}_{k=1} \left( \hat{b}^{\dagger}_k \hat{b}_{k+1} + \hat{b}^{\dagger}_{k+1} \hat{b}_{k} \right), 
\end{eqnarray}
where, we have introduced the collective system operator, $\hat{C}_N=\sum^N g_j \hat{a}_j /\mathcal{G}_N$, and the effective system-bath coupling rate, $\mathcal{G}_N =\sqrt{\sum^N_{j=1} g^2_j }$.

We next assume that the system is only weakly coupled to the bath. This is achieved by restricting that the bath-bath coupling rate $\Delta$ is much larger than the collective coupling of the system to the first bath mode, $\Delta \gg \mathcal{G}_N $. This assumption requires that the distance between the system waveguides and the first bath waveguide is large compared to the distance between each bath waveguide. By further assuming that the system and the bath are initially in a product state, $\rho(0)=\rho_{sys} (0) \otimes \rho_{bath} (0)$, starting from Hamiltonian  (\ref{HIPwithBath}), we can trace over the bath degrees of freedom to obtain the effective master equation of the system using standard techniques \cite{CarmichaelStatisticalMethods1,chinesepaper}. As the coupling between the system and the bath waveguides is weak we can make the Born-Markov approximations \cite{chinesePRA} to arrive at the following Markovian effective master equation for the system waveguides,
\begin{eqnarray}
\label{rhoCN}
\dot{\rho}_a &=& \frac{N \Gamma}{2} \left( 2 \hat{C}_N \rho_a \hat{C}^{\dagger}_N -\hat{C}^{\dagger}_N \hat{C}_N \rho_a - \rho_a \hat{C}^{\dagger}_N  \hat{C}_N  \right),
\end{eqnarray}
where, we have modelled the bath at zero temperature with, $ \rho_{bath} (0) = (|0\rangle \langle 0|)^{\otimes N}$, and introduced the effective single oscillator decay rate, $\Gamma = 2 \mathcal{G}^2_N/N \Delta$. This decay rate is a useful time scale for the system because for a system of oscillators with identical coupling rates, $g_j = g \, \forall j$, the effective single oscillator decay rate, $\Gamma = 2 N g^2/ N \Delta$, is the same as the decay rate of a single oscillator, $2 g^2/\Delta$.  In Appendix \ref{AppendixB} we discuss the effect of preparing the bath waveguides in a thermal state.

The effective master equation (\ref{rhoCN}) describes the interaction of an oscillator ensemble with a common zero temperature heat bath. It coincides with the superradiance master equation (\ref{General_SRME}) under the  the replacement, $ 2 g_i g_j / \Delta \rightarrow \gamma_{i,j}$.  We see that subject to certain assumptions  the superradiance master equation can be engineered in integrated photonics system as shown in Figure \ref{IP_Diagram}.  The three primary assumptions under which this is valid are 1) the system-system waveguide interaction is negligible ($\Omega_{i,j} \approx 0$) 2) the system waveguides only couple to the first bath waveguide ($J^{(j)}_k \approx 0$) and 3) the coupling of the system waveguides to the bath waveguide is weak ($\Delta \gg \mathcal{G}_N $).

\section{Analytic Results \label{AnalyticResults}}

In this section we calculate several experimentally observable quantities from the effective master equation (\ref{rhoCN}). These observables clearly demonstrate the effect that the common bath has on the dynamics of the oscillator ensemble. The analytic results derived in this section will be compared to the numerical solution of the Hamiltonian (\ref{Ham_AllCouplings}) in section \ref{NumericResults}. 

We begin by noting that the effective master equation (\ref{rhoCN}) describes the decay of the collective mode, defined by the operator $\hat{C}_N $, to the bath. Therefore it is useful to define three  collective operators, $\hat{M} = \sum^N_{j=1} \hat{a}^{\dagger}_j \hat{a}_j$, $\hat{R} = \hat{C}^{\dagger}_N \hat{C}_N$ and $\hat{L}= \hat{M}- \hat{R}$. Physically, $\langle \hat{M} \rangle$ is the total number of quanta in the system waveguides and $\langle \hat{R} \rangle$ is the population of the collective mode, which will decay to the bath. Experimentally $\langle \hat{M} \rangle$ can be measured by photodetectors placed at the end of the system waveguides. From the effective master equation (\ref{rhoCN}) it can be shown that the total quanta in the system evolves as,
\begin{eqnarray}\label{M}
\langle \hat{M} (t) \rangle &=& \langle \hat{L} (0)\rangle + \langle \hat{R} (0)\rangle  e^{-N \Gamma t}, 
\end{eqnarray}
where, we see that $\langle \hat{L} \rangle$ is the number of quanta that remain in the system waveguides after decay  and that the population of the collective mode decays into the bath waveguides at the rate $N \Gamma$. Due to this property we will refer to $\langle \hat{L} \rangle$ as the number of \textit{dark} quanta in the system waveguides and $\langle \hat{R} \rangle$ as the number of \textit{bright} quanta in the system waveguides. Initial states with no bright quanta,  $\langle \hat{R} (0)\rangle=0$, will not evolve under the action of the effective master equation and therefore the light will be completely trapped in the system waveguides and will not leak into the bath waveguide array.

We next consider the the defining characteristic of superradiance, the intensity of emission into the bath. The intensity into the bath waveguides can be found from the rate of change of quanta in the system waveguides,
\begin{eqnarray} \label{imp2}
I (t) &=& -\frac{\partial \langle \hat{M} (t) \rangle }{\partial t}  = N \Gamma \langle \hat{R} (0) \rangle e^{-N \Gamma t}
\end{eqnarray}
Therefore, the intensity of emission from \textit{any} state is reduced to the problem of calculating $\langle \hat{R} (0)\rangle$. Experimentally, the intensity can be found from the time series of the total system quanta $\langle \hat{M} (t) \rangle$.

To see the intensity enhancement (reduction) due to superradiance (subradiance), we separate the intensity (\ref{imp2}) into two parts, $I (t) =  I^{U} (t) + I^{C} (t)$, where the uncorrelated and correlated parts are defined, respectively,
\begin{eqnarray}
 I^{U} (t) &=&  \frac{N \Gamma}{\mathcal{G}^2_N}  e^{-N \Gamma t} \sum^N_{j=1}  g^2_j\langle  \hat{a}^{\dagger}_j (0) \hat{a}_j (0) \rangle, \label{IncohIntensity} \\
 I^{C} (t) &=& \frac{N \Gamma}{\mathcal{G}^2_N} e^{-N \Gamma t} \sum^N_{i\neq j} g_i g_j \langle \hat{a}^{\dagger}_i (0) \hat{a}_j (0) \rangle. \label{cohIntensity}
\end{eqnarray}
Normal incoherent, uncorrelated emission from an ensemble of independent oscillators with independent baths have, $\langle \hat{a}^{\dagger}_i (0) \hat{a}_j (0) \rangle = 0$ for all $i\neq j$ \cite{Agarwal1974}. On the other hand, coherent correlated emission results from the correlations within an ensemble of harmonic oscillators as they interact with a common bath, i.e. $\langle \hat{a}^{\dagger}_i (0) \hat{a}_j (0) \rangle \neq 0$ for some $i\neq j$. Therefore we associate normal radiance with the uncorrelated part, $I^{U} (t)$, and superradiance or subradiance with the correlated part, $I^{C} (t)$. Using this property we can classify initial states with a \textit{superradiance criterion}: superradiance occurs for  $I^{C} (t)>0$, normal radiance occurs when  $I^{C} (t)=0$, and subradiance occurs when $I^{C} (t)<0$. Furthermore for an initial state with total quanta $\langle \hat{M} (0)\rangle$ the maximum radiance occurs when $\langle \hat{R}(0)\rangle= \langle \hat{M}(0)\rangle$ (superradiance) and there is no emission when $\langle \hat{R}(0)\rangle= 0$ (subradiance). 

Using the superradiance criterion it is straightforward to identify three special classes of states which are superradiant, normal and subradiant, respectively. The first class we consider is the superradiant (bright) class which is defined as the Fock states of the collective mode, $|\Phi^R_B\rangle= (\hat{C}^{\dagger}_N)^R |0\rangle^{\otimes N}/\sqrt{R!}$, here the label $B$ denotes the states as bright states.  These states have no dark quanta and therefore all energy in the ensemble will decay to the bath, leaving the system in the multimode vacuum state. The second class is the product of Fock single mode states, $|n_1, n_2, \dots, n_N\rangle$, which has normal radiance as the modes are uncorrelated, $\langle \hat{a}^{\dagger}_{i} (0)\hat{a}_{j} (0)\rangle=0 \; \forall i\neq j$. The final class of states we consider are subradiant states. These states can be generated by the action of any collective operator which commutes with the collective mode, $[\hat{A}, \hat{C}^{\dagger}_N]=0$. For example a subradiant (dark) class of states can be defined as,  $|\Phi^R_D\rangle = (\hat{A}^{\dagger} )^R |0\rangle^{\otimes N} / \sqrt{R!}$, where, $\hat{A} = (g_3 \hat{a}_1 - g_1 \hat{a}_3)/\sqrt{g_1^2+ g_3^2}$ and the label $D$ denotes that the states are dark. As these states have no bright quanta,   they do not evolve under the action of the effective master equation. 

Pairwise correlations are another important observable for superradiance as these correlations are the cause of the enhanced emission intensity in superradiance. However, previous superradiance experiments on atomic ensembles have been unable to measure pairwise correlations due to the large degree of confinement of the ensemble which is necessary to engineer a common reservoir. Instead experiments probing these inter-system correlations have been restricted to observables that are averages over the entire ensemble \cite{EnsembleAverages}. However, as we describe below, it is possible to measure pairwise correlations in our proposed integrated photonics architecture.

Two-time correlation functions between two modes can be found using the the quantum regression theorem,
\begin{eqnarray}\label{TwotimeCorr}
c_{i,j} (t , 0) &\equiv& \langle \hat{b}^{\dagger}_i (t) \hat{b}_j (0) \rangle - \langle \hat{b}^{\dagger}_i (t) \rangle \langle \hat{b}_j (0) \rangle \nonumber \\
&=& c_{i,j} (0 , 0)-\frac{g_i}{ \mathcal{G}_N} (1- e^{-\frac{N \Gamma}{2} t}) \Big( \langle \hat{C}^{\dagger}_N (0) \hat{b}_j (0)  \rangle  \nonumber \\
&-&  \langle \hat{C}^{\dagger}_N (0) \rangle   \langle \hat{b}_j (0)  \rangle  \Big). 
\end{eqnarray}
We see that the correlations between the $i$-th and $j$-th waveguide build up at the rate $\frac{N \Gamma}{2}$ due to their interaction with the common reservoir. This is in contrast to normal radiance, where the oscillators decay independently  into individual reservoirs and therefore cannot build up correlations.

The three classes of states introduced above have very different two-time correlation functions. For the superradiant states, $| \Phi^R_B\rangle$, the correlation between the oscillators exponentially decays over time due to the interaction with the common bath,
\begin{equation}\label{collBasisStatecorr}
c^B_{i,j} (t , 0)  = R \frac{ g_i g_j }{\mathcal{G}^2_N} e^{-N \Gamma t},
\end{equation}
as the highly correlated initial state decays into the multimode vacuum $|0\rangle^{\otimes N}$ under the action of the effective master equation (\ref{rhoCN}). For the Fock states $|n_1, n_2, \dots, n_N\rangle $ the modes are initially uncorrelated and build up negative correlations over time,
\begin{equation} \label{corrFock}
c^N_{i,j} (t , 0)  = - \frac{ g_i g_j }{\mathcal{G}^2_N} (1- e^{-N \Gamma t}) n_j + \delta_{i,j} n_j.
\end{equation}
The finite correlation at long times occurs as the effective master equation (\ref{rhoCN}) drives the ensemble into a dark state, trapping quanta in the ensemble. Finally, subradiant states with $\langle\hat{R}(0) \rangle=0$  do not evolve over time and therefore have a constant correlation, $c_{i,j} (t , 0)=c_{i,j} (0 , 0)$. For the subradiant states $| \Phi^R_D\rangle$, we find the constant correlation functions,
\begin{equation}\label{collBasisStatecorrDARK}
c^D_{i,j} (t , 0)= c^D_{i,j} (0, 0)   = R \frac{ (g_3 \delta_{i,1}- g_1 \delta_{i,3}) (g_3 \delta_{j,1}- g_1 \delta_{j,3})}{g_1^2+ g_3^2} .
\end{equation}

As these two-time correlation functions have not previously been measured, we now detail a method for measuring, $c_{i,j} (t , 0)$ for a given initial state in the integrated photonics architecture. First, homodyne measurement can be used to perform ensemble of $n$ measurements of the $j$-th field amplitude before that field state enters the $j$-th system waveguide. We denote this ensemble of measurements as, $\{ [\hat{b}_j (0)]_1 , [\hat{b}_j (0)]_2, \dots [\hat{b}_j (0)]_{n} \}$,  where  $[\hat{b}_j (0)]_{k}$ is the outcome of the $k$-th measurement and $n$ is large. Similarly, homodyne measurement at the end of waveguide $i$ can be used to find the ensemble of $n$ measurements of the $i$-th field amplitude, $\{ [\hat{b}^{\dagger}_i (t)]_1 , [\hat{b}^{\dagger}_i (t)]_2, \dots [\hat{b}^{\dagger}_i (t)]_{n} \}$, where $t$ corresponds to the time taken to propagate the length of the waveguide. It is clear that $\langle \hat{b}_j (0) \rangle$ is the mean of the first ensemble and $\langle \hat{b}^{\dagger}_i (t) \rangle$ is the  mean of the second ensemble. Using these two ensembles it is also possible to find,
\begin{eqnarray}
\langle \hat{b}^{\dagger}_i (t) \hat{b}_j (0) \rangle &=& \frac{1}{n} \sum^n_{k=1}  [\hat{b}^{\dagger}_i (t)]_{k} [\hat{b}_j (0)]_k .
\end{eqnarray}
Therefore it is possible to measure the two-time correlation function (\ref{TwotimeCorr}) using homodyne measurement. The time interval $t$ can be varied by changing the length of the waveguides or tapping the light at different lengths.

\section{Numerical Results \label{NumericResults} }

\begin{figure}
\resizebox{0.95\columnwidth}{!}{
  \includegraphics{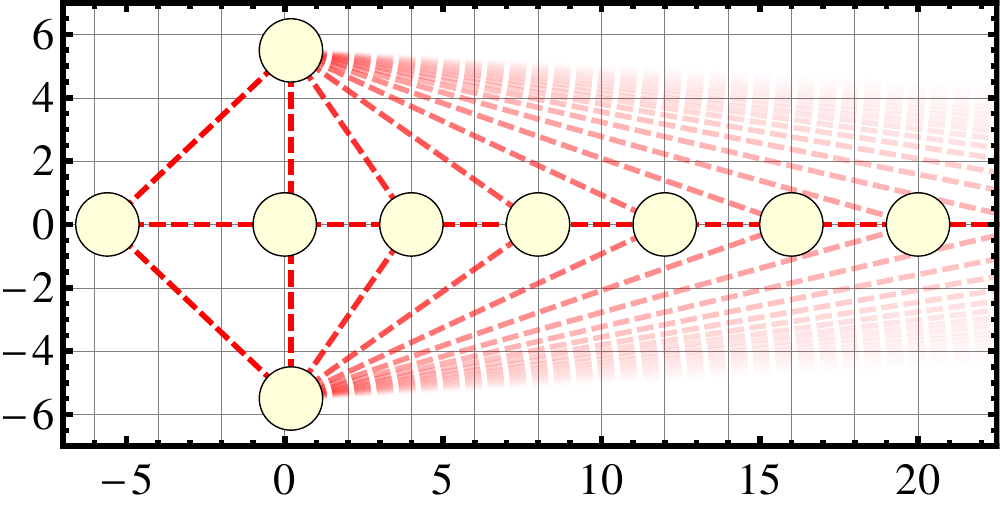}
}
\caption{Depiction of actual geometry used in the simulation. Distances between waveguides are measured in multiples of the waveguide radius. Dashed lines represent coupling rates included in the simulation.}
\label{grid_figure_WithLines}
\end{figure}

In this section we numerically solve the evolution of several initial states in a three waveguide realization of the integrated photonics system discussed in section \ref{Section_IPSystem_Intro}. These results are compared to the analytical results based on the effective master equation that were derived in section \ref{AnalyticResults}. Returning to Figure \ref{IP_Diagram} we note that there is considerable freedom to chose the physical position of the three system modes. The particular geometry we study, shown in Figure \ref{grid_figure_WithLines}, has been chosen to minimize both the system-system waveguide interaction and the coupling of system waveguides to bath waveguides other than the first bath waveguide. The coupling rates between all waveguides were calculated for the He$_{11}$ mode under the step index approximation using material parameters consistent with current direct write experiments  \cite{DirectWrite,Okamoto}.  Under these assumptions the calculated system to first bath waveguide coupling rates normalized by the bath-bath coupling rate, $\Delta$, were found to be
$(g_1, g_2, g_3)/\Delta = (0.123126,0.107251,0.123126)$. The largest coupling rates for the unwanted interactions arise from the system-system coupling rates \\ $(\Omega_{1,2},\Omega_{1,3}, \Omega_{2,3})/\Delta = ( 4.12065 \times 10^{-3}, 7.2793  \times 10^{-5}, 4.12065 \times 10^{-3})$ and the coupling to bath modes other than the first bath mode, $(J^{(j)}_2, J^{(j)}_3, J^{(j)}_4)/\Delta = (2.42439 \times 10^{-2},  5.06164  \times 10^{-4},4.89101 \times 10^{-6})$, where due to symmetry $j=1,3$. Furthermore, as the system waveguide $\hat{a}_2$ is in the same plane as the bath waveguides, we assume $J^{(2)}_k \approx 0$. The separation of time scales,  $\Delta \gg g_j \gg  \Omega_{i,j}, J^{(j)}_k$,   are consistent with the assumptions used in section \ref{Section_IPSystem_Intro}. Finally, we note that the number of bath waveguides used in the simulation was $150$, which was determined by increasing the number of bath waveguides until there was no longer any finite bath effect on the dynamics of the system waveguides.

We now consider three particular initial states that could be prepared in an experiment to observe several aspects of superradiant phenomena,
\begin{subequations}
\label{TwoPhotonStates}
\begin{eqnarray}
|\psi_{B}\rangle  &=& \frac{ (\hat{C}^{\dagger}_3 )^2}{\sqrt{2!}} |0,0,0\rangle, \\
|\psi_{N}\rangle &=& |2,0,0\rangle, \\
|\psi_{D}\rangle  &=&  \frac{ (\hat{A}^{\dagger})^2}{\sqrt{2!}} |0,0,0\rangle 
\end{eqnarray}
\end{subequations}
which we denote as bright, normal and dark, respectively. Using (\ref{imp2}) we find the intensity for each state is, $I_k (t) = 6 f_k e^{-3 \Gamma t}$, where the state dependent factors are $f_B = 1, f_N =g^2_1/\mathcal{G}^2_3, f_D=0$, with $f_B>f_N>f_D$. As expected, the superradiant state has a larger intensity than the normal state and the subradiant state does not decay, $I_D (t) =0$. Furthermore, the two-time correlation functions for these states (\ref{TwoPhotonStates}) can be found from (\ref{collBasisStatecorr})-(\ref{collBasisStatecorrDARK}) with, $R=2$ and $n_j = 2 \delta_{1,j}$. We discuss how these states can be experimentally prepared in Appendix \ref{Appendix_Initial_States}.

\begin{figure}
\subfigure[]{
\resizebox{0.93\columnwidth}{!}{
{\includegraphics{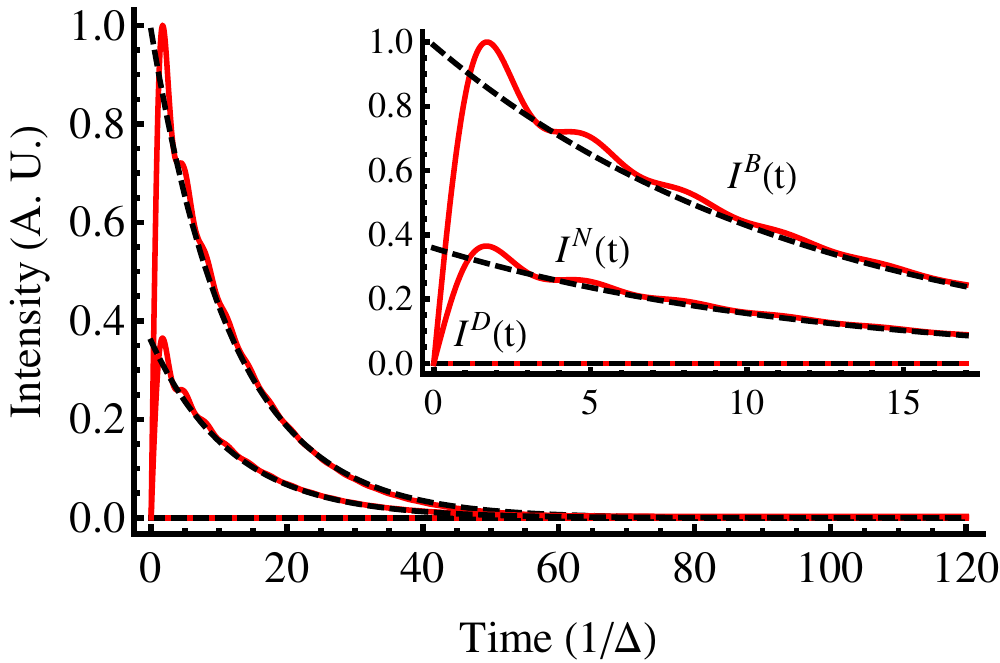}  \label{Final_IP_SRHO_Intensity}}
}
}
\subfigure[]{
\resizebox{0.96\columnwidth}{!}{
{\includegraphics{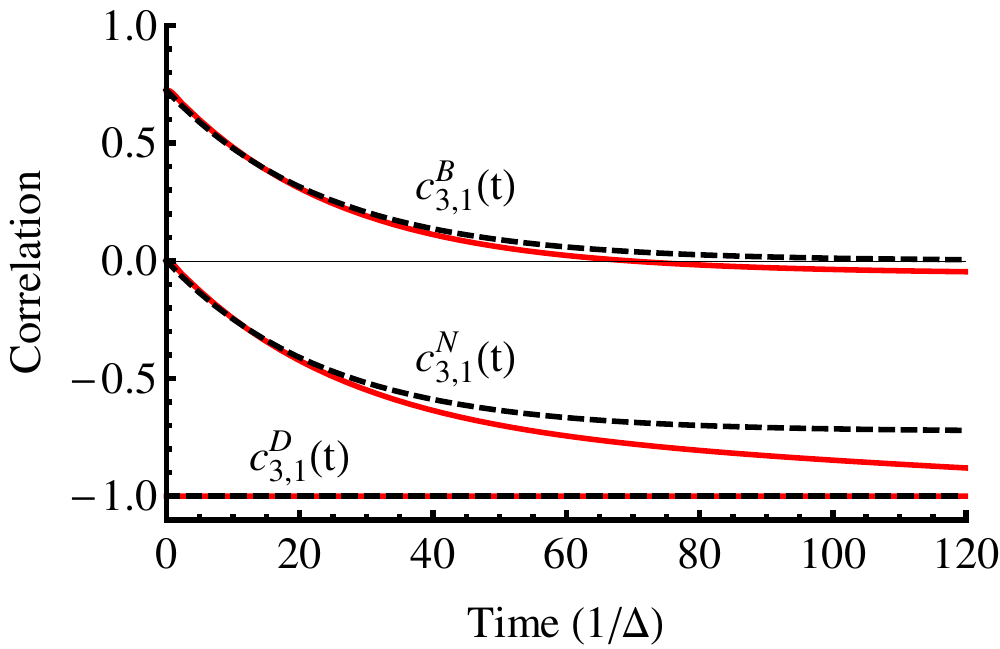}  \label{Final_IP_SRHO_Correlations}}
}
}
\caption{
Comparison of numerical solution (solid) and analytic approximation (dashed) for the two photon superradiant, normal and subradiant states (\ref{TwoPhotonStates}).
(a) Intensity of emission into the bath waveguides $I (t)$ normalized for unit peak intensity and (b) two-time correlation functions $c_{1,3} (t)$. The analytic results match the numerical results quite well, apart from   $c^B_{1,3} (t)$ and $c^D_{1,3} (t)$ which diverge due to the unwanted system-system and system-bath interactions.
}
\label{ResultsFig}
\end{figure}

The evolution of these states under the Hamiltonian (\ref{Ham_AllCouplings}) was simulated using the coupling rates calculated for the He$_{11}$ mode. The simulated intensity into the bath waveguides for the states (\ref{TwoPhotonStates}) are compared to the analytic results in Figure \ref{Final_IP_SRHO_Intensity}. Apart from minor oscillations at small times, the analytic and numerical results match very well. The oscillations are well known in these waveguide-bath systems \cite{LonghiNonexp} and are a result of  minor violations of the weak coupling assumption, $\Delta\gg \mathcal{G}_N$. The oscillations can be minimized by increasing the distance between the system and bath waveguides, however, this requires longer waveguides to observe the decay. As the length of waveguides are limited to the chip size in integrated photonics, these oscillations will always be present in some form.

The two-time correlation functions were calculated numerically from the Heisenberg equations of motion,
\begin{eqnarray}
\frac{d}{d t}\langle \hat{\mathcal{O}} (t)\rangle &=& \frac{i}{\hbar} \langle [\hat{H}, \hat{\mathcal{O}} (t) ] \rangle,
\end{eqnarray}
 using the quantum regression theorem \cite{Milburn}, where $ \hat{\mathcal{O}}= \hat{a}_j$ or $ \hat{\mathcal{O}}= \hat{b}_k$ and $\hat{H}$ is defined in (\ref{Ham_AllCouplings}). These are compared to the analytic correlation functions  (\ref{collBasisStatecorr})--(\ref{collBasisStatecorrDARK}) in Figure \ref{Final_IP_SRHO_Correlations}. The numeric and analytic correlation functions for the bright and dark states match quite well, however, the normal state has a much lower correlation than predicted. The clear difference in two-time correlations between all three states is apparent: the superradiant correlations decay exponentially, the normal correlations are initially zero and decrease exponentially due to coupling to the common reservoir and the subradiant correlations do not change over time. 

The discrepancy between the theory and simulation for the normal state is due to two interactions: the interaction of system waveguides with other system waveguides and the interaction of system waveguides with bath waveguides other than the first bath waveguide. Changing the position of the first and third system waveguides can reduce one of these interactions, however, this is at the expense of increasing the other interaction. However, as the primary aim of this paper is to show that superradiance and subradiance can be experimentally demonstrated, the discrepancy in the normal state is of little consequence. Finally, we note that the numerical correlation for the superradiant state becomes slightly negative rather than tending to zero as the analytic correlation does. This is because the unwanted interactions cause the superradiant state to decay into  a subradiant state rather than the multimode vacuum. Future work will incorporate these additional effects into the theoretical analysis of the ensemble.

\section{Conclusion \label{Conclusion}}

In summary, we have studied harmonic oscillator superradiance in an integrated photonics system. We have shown that  in this architecture harmonic oscillator superradiance and subradiance should be observable for the first time. We also developed analytic approximations for several experimental observables of the waveguide ensemble during decay. Furthermore, we demonstrated that pairwise two-time correlations can measured in this architecture. This work should be useful in future studies of collective effects and reservoir engineering in waveguide arrays.

We thank M. Steel for useful discussions. This work was supported by the Australian Research Council Centre of Excellence for Engineered Quantum Systems (CE110001013) and the European Commission projects QUANTIP \#244026 and Q-ESSENCE \#248095. S. R. would like to thank ARC (DP1094758) for financial support.

\appendix

\section{Collective Thermal Bath \label{AppendixB}}

In deriving the effective master equation (\ref{rhoCN}), we assumed that the bath modes were initially at zero temperature, $ \rho_{bath} (0) = (|0\rangle \langle 0|)^{\otimes N}$. In this section we consider preparing the bath waveguides in the initial thermal equilibrium state, $ \rho_{bath} (0) = \frac{1}{Z} e^{- \hbar \omega \sum_j  \hat{b}^{\dagger}_j \hat{b}_j /k_B T }$, where $Z$ is the partition function. Using the procedure outlined in section \ref{Section_IPSystem_Intro} it is possible to derive the effective master equation for this initial bath state,
\begin{eqnarray}\label{EffectiveMEThermal}
\dot{\rho}_a &=& \frac{N \Gamma}{2} (\bar{n}+ 1) \left( 2 \hat{C}_N \rho_a \hat{C}^{\dagger}_N -\hat{C}^{\dagger}_N \hat{C}_N \rho_a - \rho_a \hat{C}^{\dagger}_N  \hat{C}_N  \right)\nonumber\\
&+& \frac{N \Gamma}{2} \bar{n} \left( 2 \hat{C}^{\dagger}_N \rho_a \hat{C}_N -\hat{C}_N \hat{C}^{\dagger}_N \rho_a - \rho_a \hat{C}_N  \hat{C}^{\dagger}_N  \right).
\end{eqnarray}
We see that the effective master equation describes the interaction of an ensemble of oscillators with a common thermal bath.

The number of quanta in the system waveguides evolves under (\ref{EffectiveMEThermal}) as,
\begin{eqnarray}
\langle \hat{M} (t) \rangle &=& \langle \hat{L} (0)\rangle +\bar{n}+ ( \langle \hat{R} (0)\rangle - \bar{n})e^{-N \Gamma t}.
\end{eqnarray}
In contrast to the $\bar{n}=0$ case (\ref{M}), we see that there is now input of thermal quanta from the bath waveguides into the system waveguides. Furthermore, we can find the intensity from the rate of change of quanta from the collective mode into the bath,
\begin{eqnarray}\label{I2}
I (t) &=& N \Gamma \langle \hat{R} (t)\rangle \nonumber\\
&=&  N \Gamma( \langle \hat{R} (0)\rangle - \bar{n})e^{-N \Gamma t} + N \Gamma \bar{n}.
\end{eqnarray}
The primary difference from the $\bar{n}=0$ case (\ref{imp2}) is that there is now a steady state intensity from the system waveguides into the bath waveguides, $I_{ss} =  N \Gamma \bar{n}$. Similar to section \ref{AnalyticResults} we can expand the intensity (\ref{I2}) into uncorrelated and correlated parts. As the correlated part is identical to (\ref{cohIntensity}), the superradiance criterion defined in
section \ref{AnalyticResults} is applicable to this case.

\section{Initial State Preparation \label{Appendix_Initial_States}}

In this appendix we briefly describe how to prepare the superradiant and subradiant classes of initial states. Each class can be prepared by injecting a product state on a series of beam splitters. The two mode beam splitters we consider in this section perform the following transformation on the input modes \cite{BeamSplitter},
\begin{equation}
  \begin{pmatrix}
 \hat{a}'_i  \\
 \hat{a}'_j  \\
  \end{pmatrix}
= B_{i,j} (\theta)   \begin{pmatrix}
 \hat{a}_i  \\
 \hat{a}_j  \\
  \end{pmatrix}\equiv
\begin{pmatrix}
 \cos \theta & \sin \theta \\
- \sin \theta & \cos \theta \\
  \end{pmatrix}
  \begin{pmatrix}
 \hat{a}_i  \\
 \hat{a}_j  \\
  \end{pmatrix}.
\end{equation}

We first describe how to prepare the three mode subradiant class of states, $|\Phi^R_D\rangle = (\hat{A}^{\dagger} )^R |0,0,0\rangle / \sqrt{R!}$, where, $\hat{A} = (g_3 \hat{a}_1 - g_1 \hat{a}_3)/\sqrt{g_1^2+ g_3^2}$. These states can be prepared by injecting the product state, $|R,0,0\rangle$, onto a beam splitter acting on the first and third modes. The required transformation is achieved by the beam splitter defined by $B_{3,1} (\theta_D)$, with 
$\theta_D= \sin^{-1} (g_1/\sqrt{g_1^2+ g_3^2})$.

We now describe how to use two beam splitters to prepare the three mode superradiant class of states, $|\Phi^R_B\rangle= (\hat{C}^{\dagger}_3)^R |0\rangle^{\otimes N}/\sqrt{R!}$, where we recall $\hat{C}_3 = (g_1 \hat{a}_1+g_2 \hat{a}_2+g_3 \hat{a}_3 )/\mathcal{G}_3$. If we inject the state, $|R,0,0\rangle$, onto a beam splitter acting on the first and second modes, $B_{1,2} (\theta_B)$, we place these two modes in a superposition with the third in the vacuum state. We next use a beam splitter on the second and third modes, $B_{2,3} (\varphi_B)$. The superradiant class of states can then be prepared by choosing, $\theta_B= \cos^{-1} (g_1/\mathcal{G}_3)$ and $\varphi_B =  \sin^{-1}  (g_3/ \sqrt{g_2^2+ g_3^2})$. 

We finally note that $N$ mode classes of superradiant and subradiant states can be prepared in a similar way with additional beam splitters. In particular, the $N$ mode superradiant states $(\hat{C}^{\dagger}_N)^R |0\rangle^{\otimes N}/\sqrt{R!}$ can be prepared by injecting the state, $|R\rangle |0\rangle^{\otimes (N-1)}$, onto $N-1$, beam splitters.

\end{document}